\documentclass[a4paper]{article}

\usepackage{icrc2013}
\usepackage[english]{babel}

\title{Comparison of LOPES data and CoREAS simulations using a full detector simulation}

\newcommand{\etal}{\MakeLowercase{\textit{et al. }}} % "et al."
\shorttitle{K. Link \etal Comparison of LOPES and CoREAS}

\authors{K.~Link$^{1}$, W.D.~Apel$^{2}$, J.C.~Arteaga-Vel\'azquez$^{3}$, L.~B\"ahren$^{4}$, K.~Bekk$^{2}$, M.~Bertaina$^{5}$, P.L.~Biermann$^{6,2}$, J.~Bl\"umer$^{1,2}$, H.~Bozdog$^{2}$, I.M.~Brancus$^{7}$, E.~Cantoni$^{5,8}$, A.~Chiavassa$^{5}$, K.~Daumiller$^{2}$, V.~de~Souza$^{9}$, F.~Di~Pierro$^{5}$, P.~Doll$^{2}$, R.~Engel$^{2}$, H.~Falcke$^{4,10,6}$, B.~Fuchs$^{1}$, D.~Fuhrmann$^{11}$, H.~Gemmeke$^{12}$, C.~Grupen$^{13}$, A.~Haungs$^{2}$, D.~Heck$^{2}$, J.R.~H\"orandel$^{4}$, A.~Horneffer$^{6}$, D.~Huber$^{1}$, T.~Huege$^{2}$, P.G.~Isar$^{14}$, K-H.~Kampert$^{11}$, D.~Kang$^{1}$, O.~Kr\"omer$^{12}$, J.~Kuijpers$^{4}$,  P.~{\L}uczak$^{15}$, M.~Ludwig$^{1}$, H.J.~Mathes$^{2}$, M.~Melissas$^{1}$, C.~Morello$^{8}$, J.~Oehlschl\"ager$^{2}$, N.~Palmieri$^{1}$, T.~Pierog$^{2}$, J.~Rautenberg$^{11}$, H.~Rebel$^{2}$, M.~Roth$^{2}$, C.~R\"uhle$^{12}$, A.~Saftoiu$^{7}$, H.~Schieler$^{2}$, A.~Schmidt$^{12}$, F.G.~Schr\"oder$^{2}$, O.~Sima$^{16}$, G.~Toma$^{7}$, G.C.~Trinchero$^{8}$, A.~Weindl$^{2}$, J.~Wochele$^{2}$, J.~Zabierowski$^{15}$, J.A.~Zensus$^{6}$\\
- LOPES Collaboration -}

\afiliations{
$^1$Institut f\"ur Experimentelle Kernphysik, Karlsruher Institut f\"ur Technologie (KIT), Germany\\
$^2$Institut f\"ur Kernphysik, Karlsruher Institut f\"ur Technologie (KIT), Germany\\
$^3$Universidad Michoacana, Instituto de F\'{\i}sica y Matem\'aticas, Morelia, Mexico\\
$^4$Department of Astrophysics, Radboud University Nijmegen, The Netherlands\\
$^5$Dipartimento di Fisica dell' Universit\`a Torino, Italy\\
$^6$Max-Planck-Institut f\"ur Radioastronomie Bonn, Germany\\
$^7$National Institute of Physics and Nuclear Engineering, Bucharest, Romania\\
$^8$Osservatorio Astrofisico di Torino, INAF Torino, Italy\\
$^9$Universidade S$\tilde{a}$o Paulo, Instituto de F\'{\i}sica de S$\tilde{a}$o Carlos, Brasil\\
$^{10}$ASTRON, Dwingeloo, The Netherlands\\
$^{11}$Fachbereich Physik, Universit\"at Wuppertal, Germany\\
$^{12}$Institut f\"ur Prozessdatenverarbeitung und Elektronik, Karlsruhe Institute of Technology (KIT), Germany\\
$^{13}$Fachbereich Physik, Universit\"at Siegen, Germany\\ $^{14}$Institute of Space Science, Bucharest, Romania\\
$^{15}$National Centre for Nuclear Research, Department of Cosmic Ray Physics, {\L}\'od\'z, Poland\\
$^{16}$Department of Physics, University of Bucharest, Bucharest, Romania}

\email{katrin.link@kit.edu}

\abstract{The LOPES experiment at the Karlsruhe Institute of Technology, Germany, has been measuring radio emission of air showers for almost 10 years. For a better understanding of the emission process a detailed comparison of data with simulations is necessary. This is possible using a newly developed detector simulation including all LOPES detector components. After propagating a simulated event through this full detector simulation a standard LOPES like event file is written. LOPES data and CoREAS simulations can then be treated equally and the same analysis software can be applied to both. This gives the opportunity to compare data and simulations directly. Furthermore, the standard analysis software can be used with simulations which provide the possibility to check the accuracy regarding reconstruction of air shower parameters. We point out the advantages and present first results using such a full LOPES detector simulation. A comparison of LOPES data and the Monte Carlo code CoREAS based on an analysis using this detector simulation is shown.}


\begin{document}
\maketitle

\section{Introduction}
Cosmic ray air showers produce radio emission while traveling through the Earth's magnetic field. This radio emission is measured by the LOPES experiment \cite{FalckeNature2005} at the Karlsruhe Institute of Technology (KIT), Germany. Besides a trigger on high-energy events LOPES profits from detailed shower parameters provided by the co-located particle detector KASCADE-Grande \cite{AntoniApelBadea2003,ApelArteagaBadea2010}. LOPES was measuring up to the beginning of 2013 using various experimental setups: Starting in 2003 with 10 east-west aligned antennas (LOPES 10) first the number of antennas was increased to 30 (LOPES 30) and then half of them were rotated by 90$^{\circ}$ (LOPES 30 pol). Finally the antenna type changed from an inverted-v-shaped dipole to a tripole antenna (LOPES 3D) \cite{huberpaper} to measure the complete three-dimensional electric field vector. All setups used a frequency range of \mbox{40-80$\,$MHz} and a sampling rate of 80$\,$MHz. With LOPES as an array of antennas with very good time synchronization, an interferometric analysis is possible which allows us to do air shower measurements even in the noisy environment at KIT.

\section{Electric field}

The measurement of the electric field with an antenna is mathematically a projection of the vector on the antenna plane which is described by a dot product of the electric field vector $\vec{E}$ and the gain vector $\vec{G}$. This leads to the following equation:

\begin{equation}
S_{\mbox{ant}} = \vec{E}\cdot\vec{G}
\label{equ:gainscalar}
\end{equation}

The gain vector $\vec{G}$ describes the sensitivity for a particular antenna, see subsection \ref{sec:gain}. The resulting scalar S$_{\mbox{ant}}$ corresponds to the measured voltage at the antenna foot-point. The electric field reduces to a two-dimensional vector in the shower plane due to the fact that the radio emission from air showers is a transverse wave. Therefore, with a known arrival direction, at least two measurements are necessary for a correct reconstruction. For the LOPES 10 and LOPES 30 setup only one antenna is installed at each position and even for the LOPES 30 pol setup only at five positions two antennas are installed. Therefore it is generally not possible to do a correct reconstruction of the electric field vector with LOPES data. In the standard LOPES analysis, a simplified reconstruction is done using the assumption that the field vector is orientated along the antenna which corresponds to

\begin{eqnarray}
S_{\mbox{ant}} &=& |\vec{E}|_{\mbox{ant}}\cdot | \vec{G}_{\mbox{ant}}| \nonumber \\
\Rightarrow E_{\mbox{ant}} &=& \frac{S_{\mbox{ant}}}{{G}_{\mbox{tot,ant}}}
\label{equ:simplegain}
\end{eqnarray}  

For the LOPES 3D setup the correct reconstruction was implemented and can also be used for the five stations with two antennas of the LOPES 30 pol setup, for details see \cite{Huber2013}. With the correct reconstruction it is possible to do a comparison of data and simulation by comparing the reconstructed electric field vector of measurements with the simulated electric field vector. 
For the simplified reconstruction the reconstructed component of the electric field does not correspond directly to a component of the true electric field vector. To calculate the expected signal at the antenna foot-point a detector simulation is needed which is in principle represented by equation \ref{equ:gainscalar}. Additional all hardware and software components of the LOPES detector are included. Using such a detector simulation it is possible to do a comparison of data and simulations using directly the voltage at the antenna foot-point or applying the simplified reconstruction on both. This allows us also to test the precision of the standard analysis pipeline with simulations or to investigate the influence of detector effects like quantization or sampling rate and to analyze the influence of noise.  

\subsection{Antenna gain simulations}
\label{sec:gain}

For all comparisons of data and simulations the sensitivity, respectively the gain, of the antenna is important. For the LOPES antennas two simulations are available, one including a metal pedestal and one without. To have a defined ground it was decided to build the LOPES antennas above such a pedestal. Former studies showed already, that the influence of this pedestal is probably overestimated \cite{pedestalStudies}. The main difference in the two antenna simulation affect the phases of the antennas which are only used for the detector simulations. Therefore, it was now reasonable to switch completely to the simulations of the antenna without the pedestal. For future studies a simulation including a more realistic pedestal is planned. 
 
\section{Simulations}

\begin{figure}[!h]
  \centering
  \includegraphics[angle=270, width=0.43\textwidth]{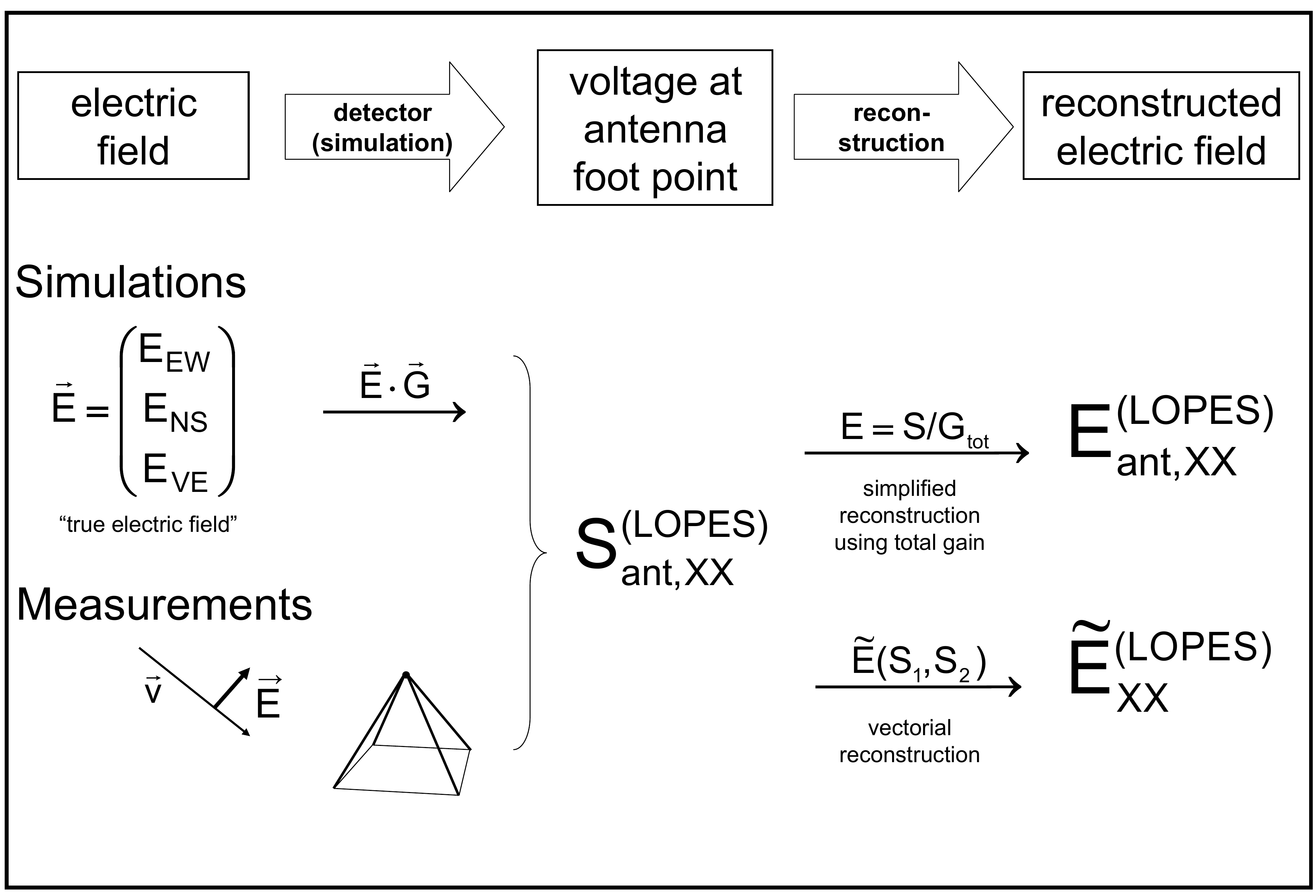}
  \caption{Schematic representation of the available quantities for the electric field. XX stands for EW, NS or VE.}
  \label{fig:simchart}
 \end{figure}

To understand the origin of the radio emission and the involved emission processes, a detailed simulation of the emission physics is necessary. Two simulation codes for radio emission of air showers are the Monte Carlo code CoREAS \cite{HuegeCoreas} and REAS3 \cite{REAS3}. REAS3 is based on the endpoint formalism and uses histogrammed air shower information of CORSIKA \cite{Corsika}. It is a parameter-free simulation including the refractive index in version REAS3.11. Implemented directly in CORSIKA, CoREAS is also based on the endpoint formalism and a parameter-free simulation including the whole complexity of an air shower. It is possible to simulate the radio emission of a particular air shower measured with LOPES taking into account the parameters given by KASCADE-Grande, i.e. energy and arrival direction. The simulation provides the full time-dependence of the electric field vector at a given position, so for each antenna of the LOPES array the electric field is known. With the detector simulation the signal at the antenna foot-point is calculated and also the simplified reconstruction is done. Figure \ref{fig:simchart} gives an overview of the available quantities that are analogous to the electric field.

\section{Impact of simplified reconstruction}

\begin{figure}[!th]
  \centering
  \includegraphics[angle=90, width=0.45\textwidth]{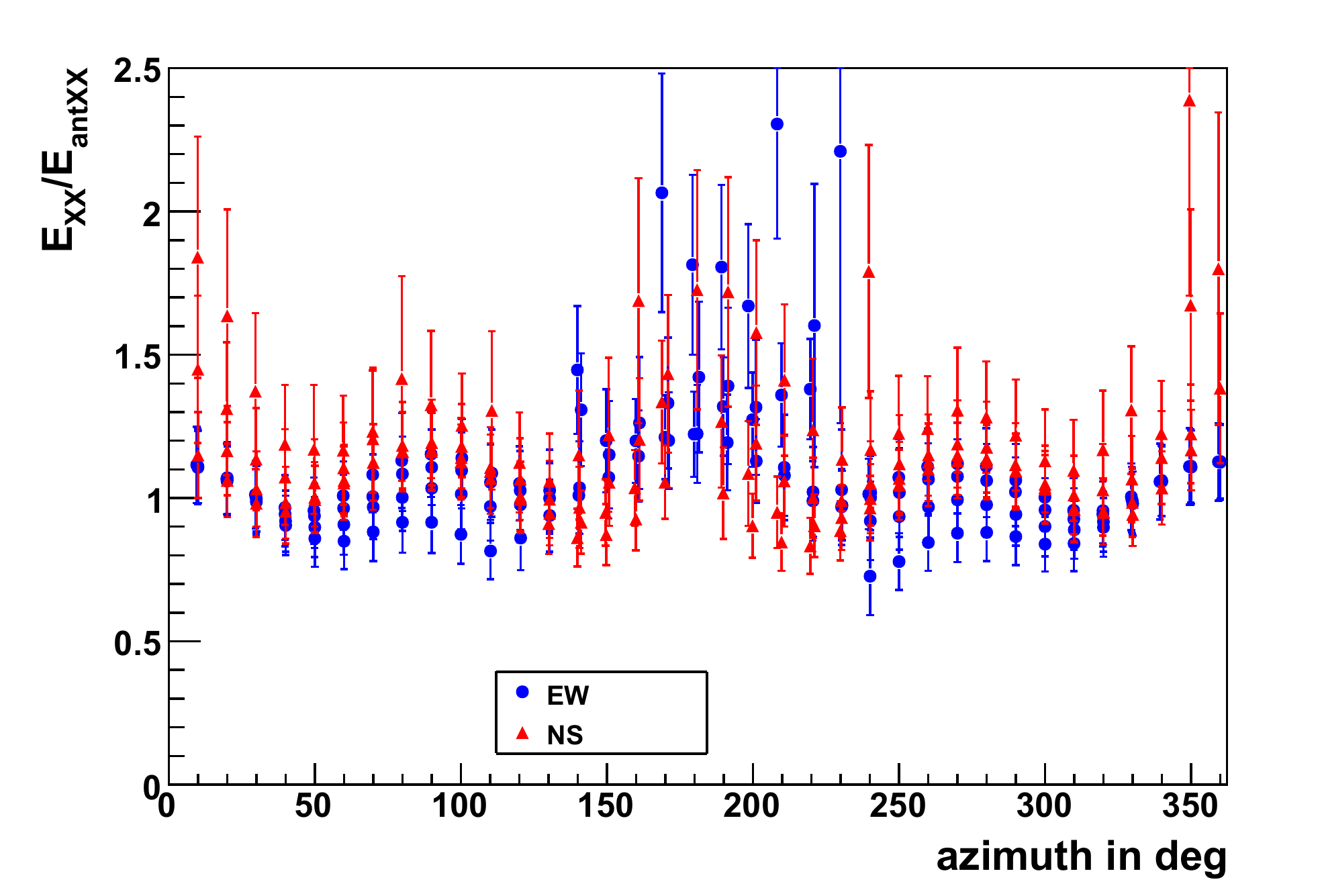}
  \includegraphics[angle=90, width=0.45\textwidth]{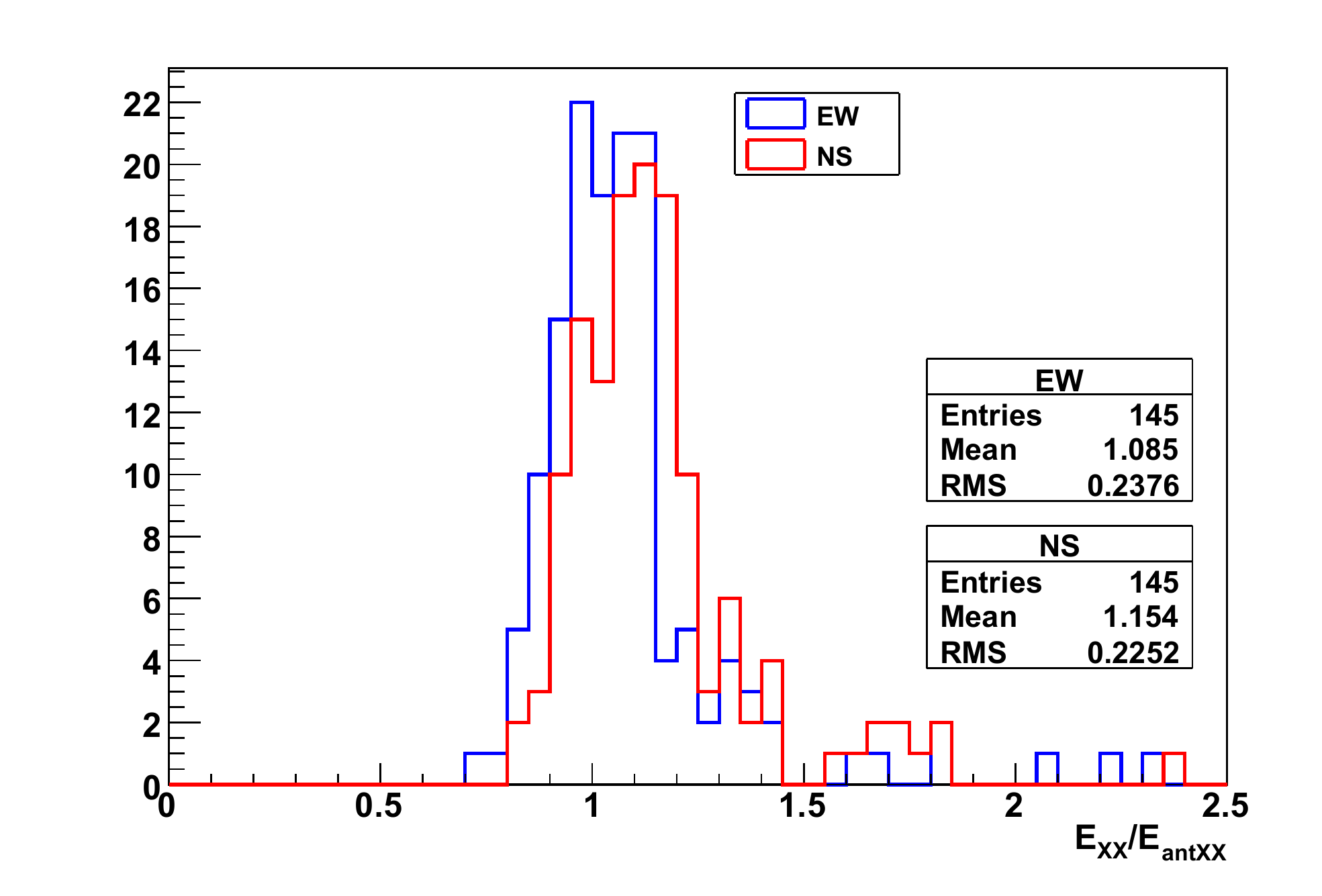}
  \caption{Ratio of the true electric field and the simplified reconstructed electric field for a library of 145 events simulated with REAS3. }
  \label{fig:librarycomp}
 \end{figure}

In former comparisons of data and simulations, the simplified reconstructed electric field $E_{antXX}^{LOPES}$ and the simulated component of the electric field $E_{XX}$ were studied, with XX either be east-west (EW), north-south (NS) or vertical (VE). To quantify how the simplification influences the reconstruction, a library of 145 events was simulated with REAS3, covering different arrival directions (36 different azimuth angles and 5 different zenith angles) for an energy of $10^{17}\, $eV. For a comparison of the true with the reconstructed electric field an exponential function

\begin{eqnarray}
\epsilon(d)=\epsilon_{100}\cdot exp\left((d-100\mbox{m})/R_{0}\right)
\label{equ:exponential}
\end{eqnarray}
 
is fitted to the lateral distribution, with the fitparameters $\epsilon_{100}$, which is the amplitude at 100m, and the slope parameter $R_{0}$. In figure \ref{fig:librarycomp} the deviation between the reconstructed electric field $E_{antXX}\propto \epsilon_{100,antXX}$ and true electric field component $E_{XX}\propto \epsilon_{100,XX}$ is shown. There is an angle-dependent deviation visible for both the east-west and the north-south component.

\begin{figure}[!t]
  \centering
  \includegraphics[angle=90, width=0.45\textwidth]{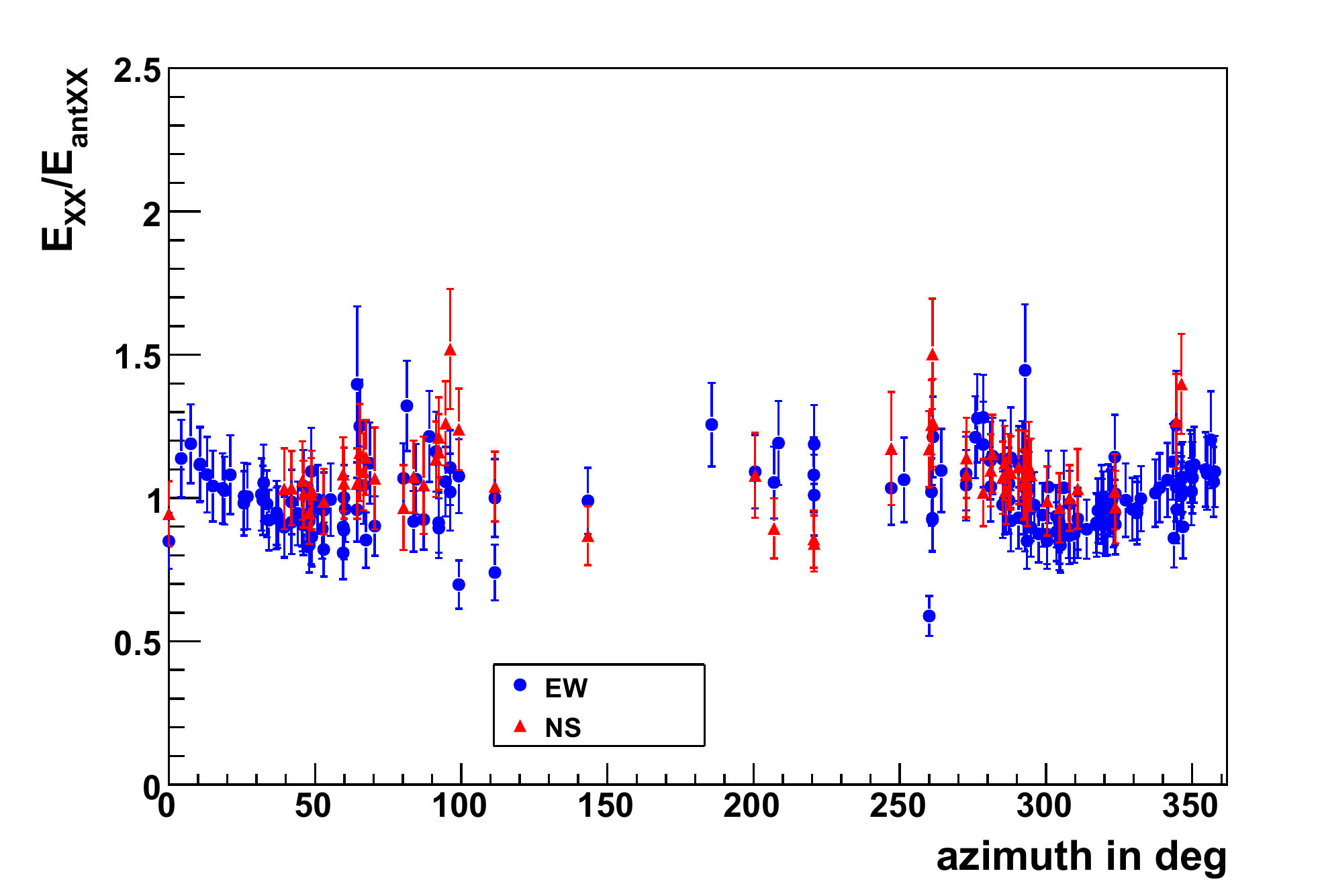}
  \includegraphics[angle=90, width=0.45\textwidth]{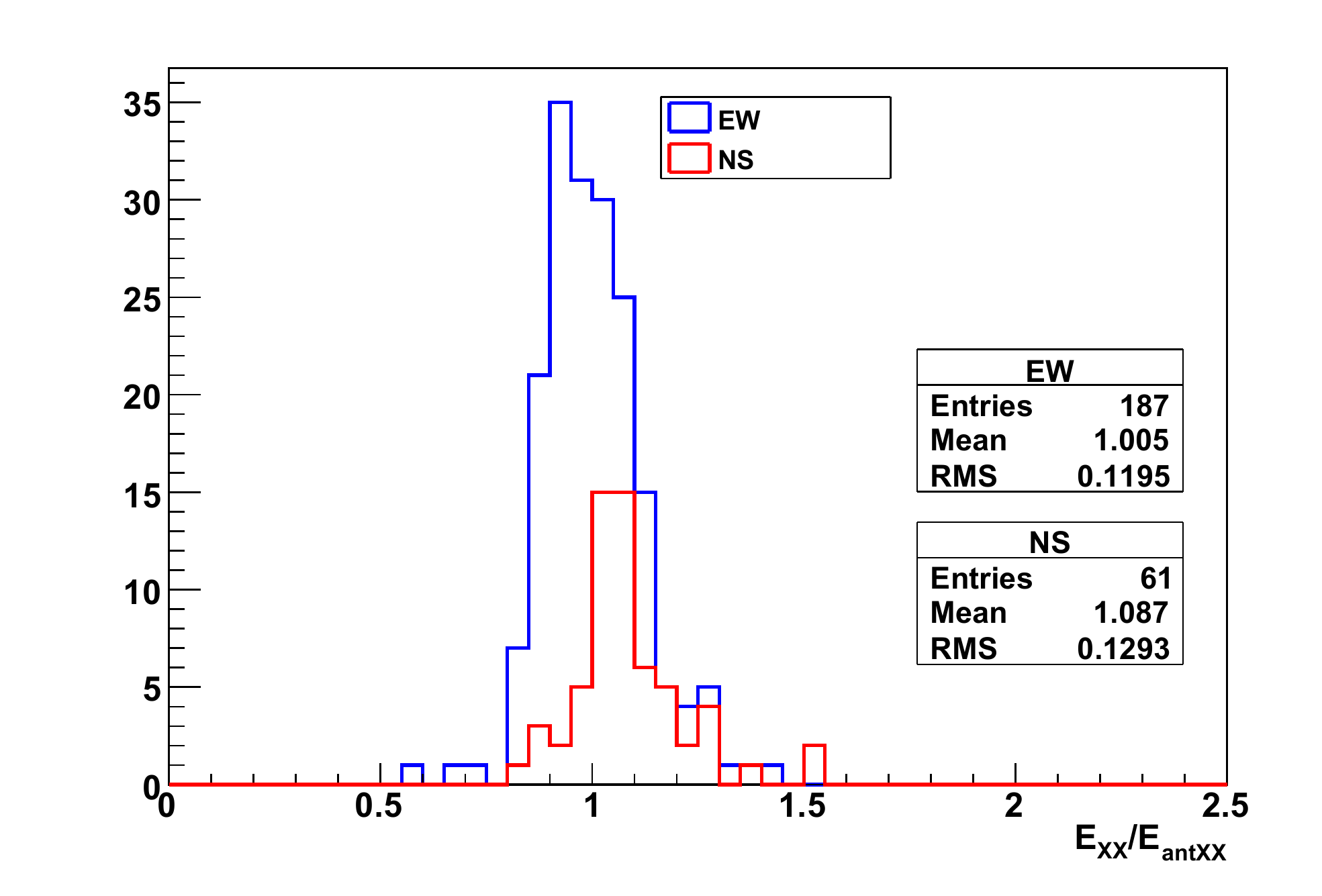}
  \caption{Ratio of the true electric field and the simplified reconstructed electric field for events simulated with CoREAS corresponding to events measured with LOPES.}
  \label{fig:LOPESselimprec}
 \end{figure}

For showers coming from the south, which is from $\sim180^{\circ}$, a big difference is visible. For the NS-component a big difference is also visible for showers coming from the north, i.e. $\sim0^{\circ}$. For single events this can be more than 50\%. On average the reconstructed electric field is bigger than the true electric field, especially for the NS-component. But it is necessary to treat events with small signals with care: No noise is added to the simulations, this means the signal strength can be much smaller than the normal noise level. Therefore artifacts can occur which will never be seen in data, e.g. for the quantization of the ADC. Especially for showers with a geometry leading to high deviation (south and north) also small signals are predicted. To get rid of this problem a cut can be applied on $eps_{100,XX}$. This leads to much smaller deviations and also the mean is closer to zero. To investigate the influence also for events measured with LOPES a set of around 200 events measured by LOPES was simulated with CoREAS using the air shower information given by KASCADE. These events were selected to have a good signal-to-noise ratio in the LOPES data, either for north-south or east-west antennas. Unlike for the REAS3 simulations used in figure \ref{fig:librarycomp}, the realistic refractive index is now included. The comparison of the true and simplified electric field components is shown in figure \ref{fig:LOPESselimprec}. Since LOPES only measures events with a signal above noise events with a geometry leading to small signals in both antennas, EW and NS, are not in this selection. However this selection includes events with a high signal in one antenna direction and a small, not visible signal, in the other one. Therefore a cut on $eps_{100,XX}>1.5\,\mu V/m/MHz$ is applied to avoid the artifacts mentioned above. Additionally the fit of the lateral distribution must not fail. Especially for the north-south antennas most of the events do not pass the cut on the $eps_{100,XX}$, so only few events remain. The mean deviation is much smaller than for the library events because the events with small predicted signals also show the most extreme deviation and these events are not included because of the detection threshold of LOPES.

\section{Comparison of data and simulation}

\begin{figure}[!th]
  \centering
  \includegraphics[angle=90, width=0.45\textwidth]{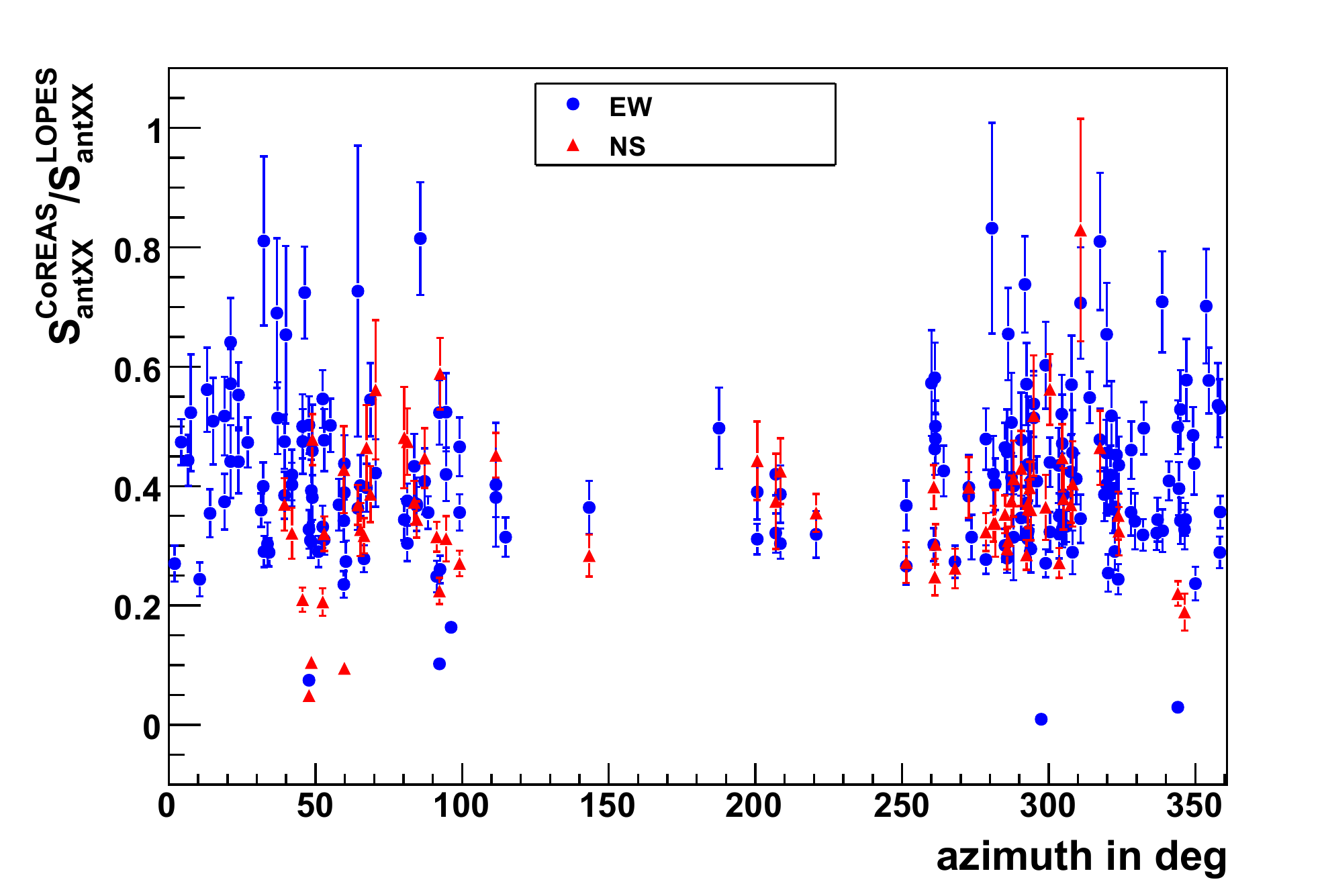}
  \includegraphics[angle=90, width=0.45\textwidth]{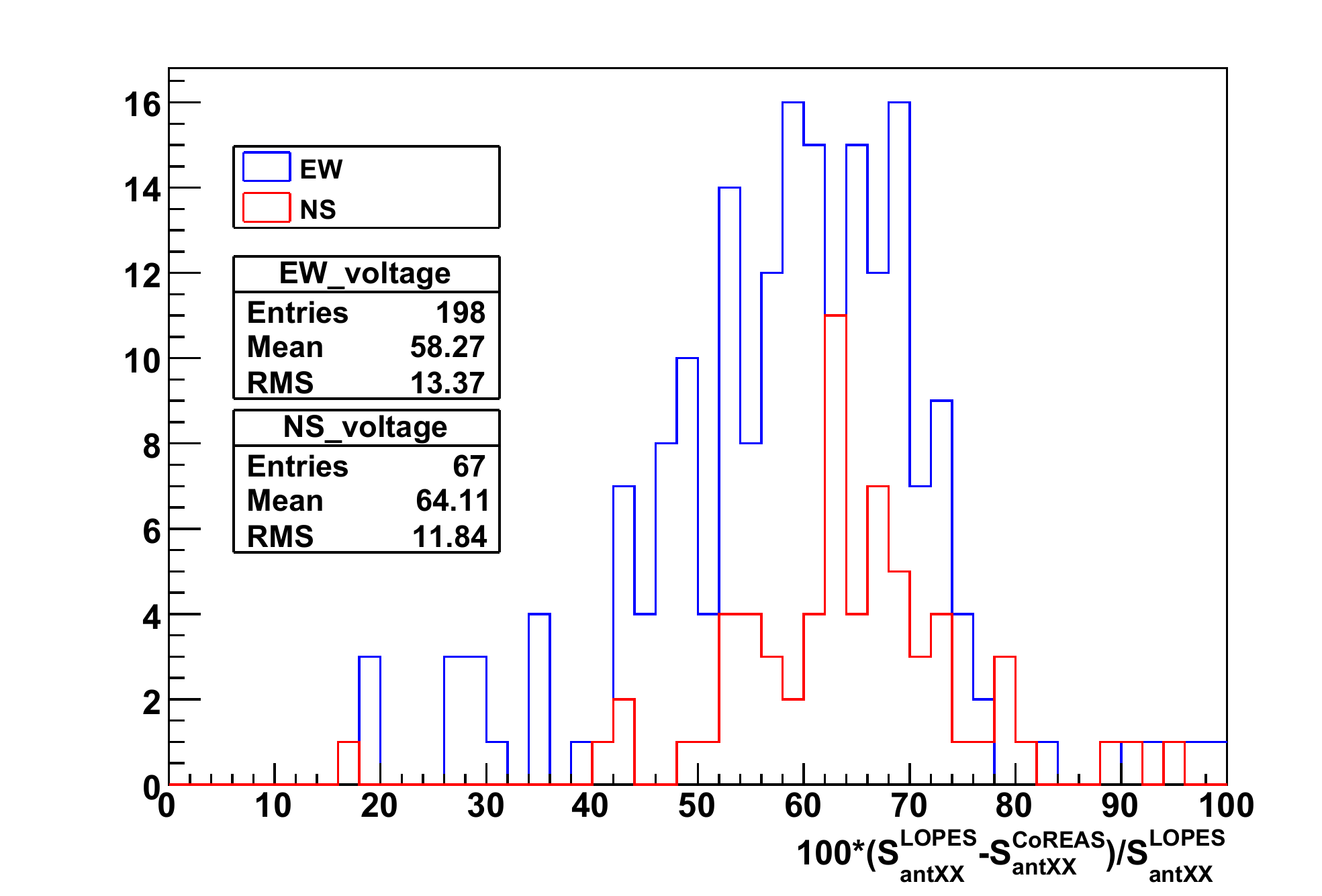}
  \caption{Top: Ratio of LOPES and CoREAS $\epsilon_{100}$ values based on the voltage at the antenna foot-point $S_{antXX}$. Bottom: Relative deviation of LOPES and CoREAS $\epsilon_{100}$ values based on the voltage at the antenna foot-point $S_{antXX}$.}
  \label{fig:antfootpoint}
 \end{figure}

Using the full detector simulation it is possible to compare CoREAS or REAS simulations and LOPES data based on the same reconstructed quantities. Either the voltage at the antenna foot-point or the output of the simplified reconstruction can be used. For a comparison based on the antenna foot-point, an exponential function, like in equation \ref{equ:exponential}, is fitted to the lateral distribution of the ADC voltage of each antenna. In figure \ref{fig:antfootpoint} the amplitude parameter $\epsilon_{100,voltage}$ of CoREAS simulations is compared to $\epsilon_{100,voltage}$ of LOPES data. On average the voltage of LOPES is more than two times higher then the predicted voltage of CoREAS. This was already seen in other comparisons of LOPES data and CoREAS simulations, see for example \cite{FrankICRC}. 

\begin{figure}[!th]
  \centering
  \includegraphics[angle=90, width=0.45\textwidth]{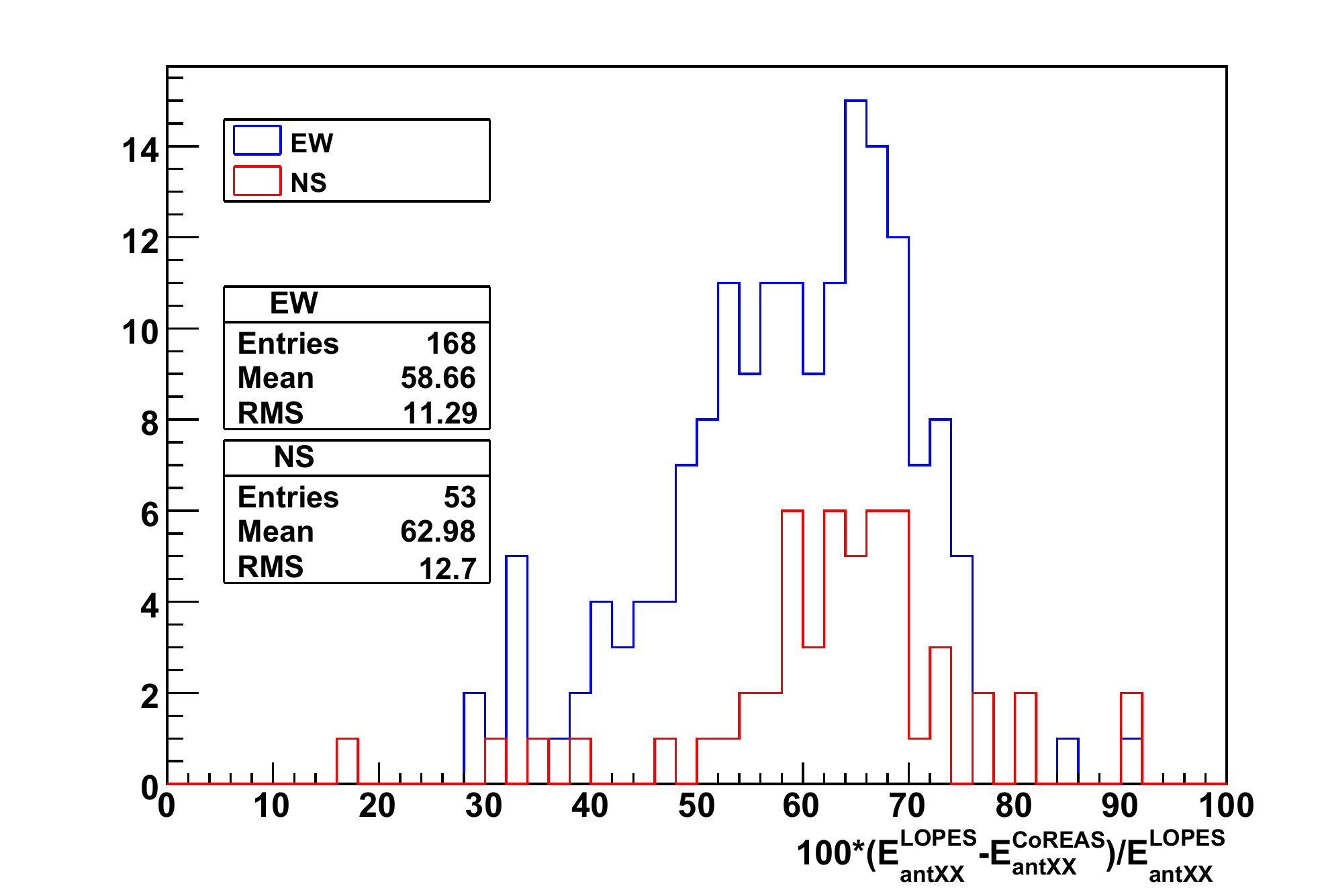}
  \caption{Relative deviation of LOPES and CoREAS $\epsilon_{100}$ values based on the simplified reconstructed electric field $E_{antXX}$.}
  \label{fig:histograms}
 \end{figure}

For a better comparability with other analyses also histograms of the $\epsilon_{100}$ values based on the simplified reconstructed electric field are shown in figure \ref{fig:histograms}. Since the simplified reconstruction corresponds only to a multiplication with the same constant factor for all antennas for each event the same deviation is visible. The scatter and the mean values are comparable with the results in reference \cite{FrankICRC}. 

\begin{figure}[!th]
  \centering
  \includegraphics[angle=90, width=0.45\textwidth]{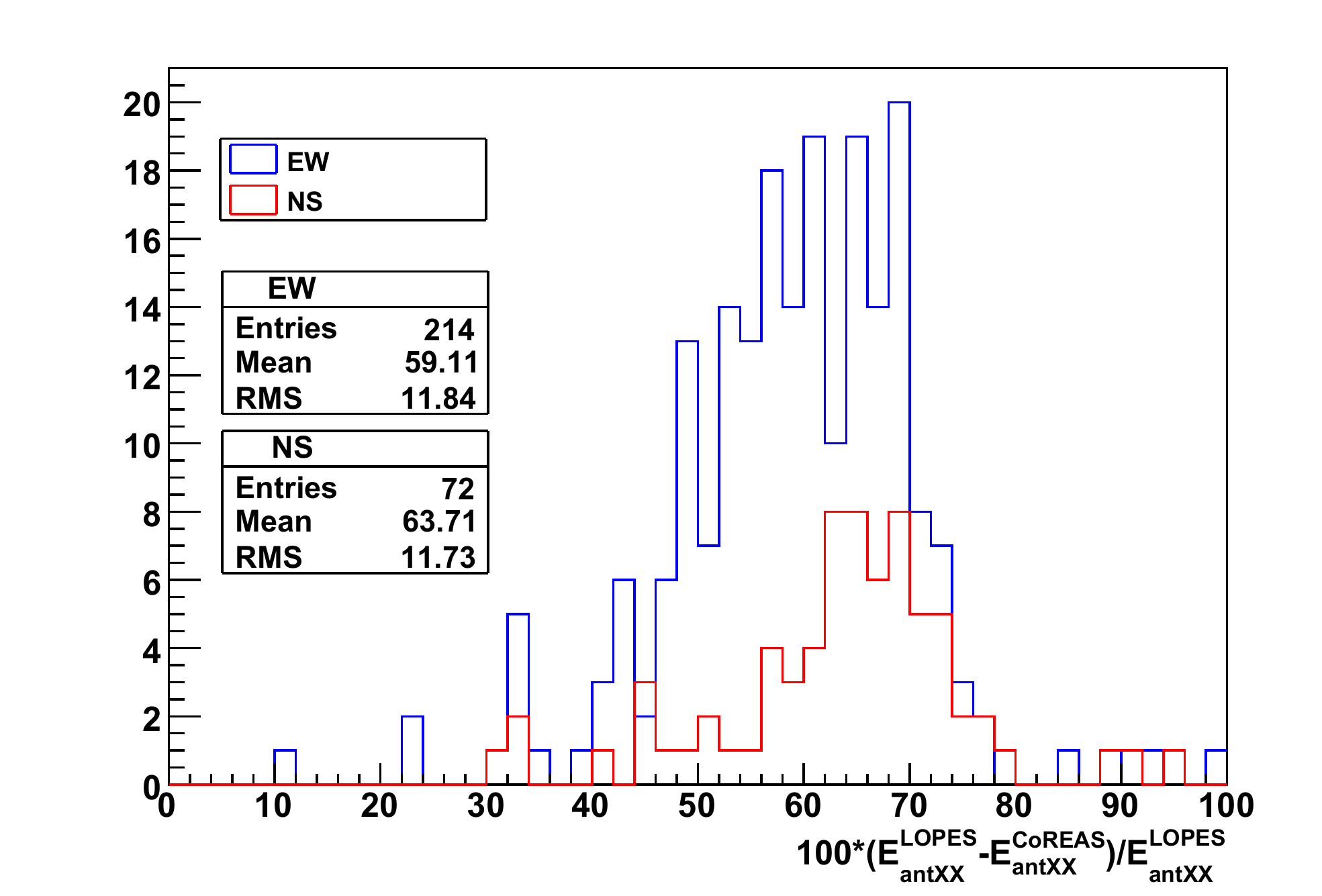}
  \caption{Relative deviation of LOPES and CoREAS amplitude based on the cross-correlation-beam calculated from the simplified reconstructed electric field.}
  \label{fig:ccheight}
 \end{figure}

Since the $\epsilon_{100}$ values depend on the fitting function of the lateral distribution this value is influenced by the fitting routine. Using instead the cross-correlation beam \mbox{(CC-beam)}, which is calculated in the standard LOPES analysis, we do not have to make any assumptions on the lateral distribution function. Figure \ref{fig:ccheight} shows a comparison of LOPES data and CoREAS simulations based on this CC-beam calculated from the simplified reconstructed electric field.

A cut on a successful fit of the lateral distribution is not needed anymore, hence more events can be used for the analysis. The RMS and the mean of the distribution are similar to the those of the $\epsilon_{100}$ values. Taking into account the scale uncertainties of 35\% for the LOPES amplitude and 20\% scale uncertainty for the energy reconstruction of KASCADE it might be still possible that CoREAS simulations can describe the LOPES data.

\section{Conclusion}

Using a full detector simulation, a new way of comparing data and simulations is possible. Instead of a comparison of the reconstructed electric field, obtained by a simplification, with the true electric field components, a comparison based on the voltage at the antenna foot-point is possible. We were able to show that the simplification introduces a spread of around 12\% for the reconstruction of the electric field components for events measured with LOPES. An average difference is only visible taking into account the full azimuthal range. For events that can be measured with LOPES the mean of the simplified electric field is almost the same as for the true electric field. A comparison of the true electric field component and the simplified reconstructed electric field for LOPES is thus warrantable. For the comparison of CoREAS and LOPES still an average difference of around a factor of two is existing. A more sophisticated analysis on the uncertainties is needed to finally exclude or confirm the predicted amplitudes of CoREAS simulations: For this analysis only simulations for iron induced air showers were considered and no noise was added to simulations. Furthermore the gain of the antennas does not include the metal pedestal at all. For a final comparison also these aspects need to be investigated.

\vspace*{0.5cm} \footnotesize{
{\bf Acknowledgement:} 

{This research has been supported by grant number VH-NG-413 of the Helmholtz Association.}}

\end{document}